\documentstyle[12pt,a4]{article}

\def\<{\langle}
\def\>{\rangle}
\def\/{\sqrt{}}

\def\d{\delta}
\def\l{\lambda}

\begin{document}

\vskip 1.0 cm 
\centerline{\Large 
Spectral Representation and 
the Averaging Problem in Cosmology
\footnote{
To appear in {\it General Relativity and Gravitation}. \\
This work is the development  of  a talk given at South African Relativistic 
Cosmology Conference, February 1-5 1999, 
Department of Mathematics and Applied Mathematics, University 
of Cape Town, South Africa.\\
This work has been financially supported by  Inamori Foundation, Japan.
 }}
\vskip .8 cm 

\centerline{\large {\sc Masafumi Seriu}\footnote{
Institute of Cosmology, Department of Physics \& Astronomy, 
Tufts University, Medford, MA 02155, USA;\ \  Department of Physics, 
Fukui University, Fukui 910-8507, Japan (permanent address).}}

\vskip 1.0cm

\begin{abstract}
We investigate   the averaging problem in cosmology as  the 
problem of  introducing a distance between spaces. 

We first  introduce the spectral distance,  
 which is a measure of closeness between spaces 
defined in terms of  the spectra of the Laplacian. 
Then we define  ${\cal S}_N$, a space of all spaces equipped with 
the spectral distance. 
We argue that ${\cal S}_N$ can be regarded as a metric space and 
that it also possess  other desirable properties. 
These facts make ${\cal S}_N$  a suitable  arena for spacetime physics.

  We apply the spectral framework to  the averaging problem:  
     We   sketch the model-fitting procedure 
     in terms of  the spectral representation, and  also 
     discuss briefly 
      how to analyze the dynamical aspects of the averaging procedure 
     with  this scheme.  
     In these analyses, we are  naturally led to the concept of  
     the apparatus- and the scale-dependent effective evolution of 
     the universe. 
     These observations suggest  that 
     the spectral scheme  seems to be suitable for the quantitative 
     analysis of the averaging problem in cosmology.   
\end{abstract}

\vskip 1cm
\section{Introduction}

The {\it averaging problem} is one of the fundamental problems   
in cosmology  that we have not yet understood sufficiently 
so far\cite{AVE,AVE2}.

It can be summarized as follows: In cosmology we want to 
understand the whole picture of our universe. However since 
the structure of the universe is so  complicated that we can understand it 
only with the help of some format of recognition, viz.  cosmological 
models.  Thus, cosmology in principle requires a mapping procedure 
from reality to a model. Alternatively, one may regard this procedure 
as the procedure of averaging the  real, complicated geometry in some manner 
in order to assign 
a simplified model-geometry to it.   
Now the problem is that Einstein equation is 
highly nonlinear, so that  the effective dynamics of 
 the  averaged spatial geometry is expected to be  highly complicated. 
Moreover, the effective dynamics of the averaged spatial geometry 
would not in general match the dynamics of the assigned model. 
Hence we should first analyze and understand the averaging procedure 
{\it itself}, otherwise 
we would make serious mistakes in finding out  the evolution of the 
universe and/or matter content of the universe. 
If we state symbolically,
the averaging procedure and  the Einstein equation  
do not commute with  each other.

Once we start  investigating  the averaging procedure itself, 
we immediately face with the trouble that we do not have a suitable 
`language' for describing it. In order to formulate the approximation  
of the real geometry 
 by a certain model, the concepts like  `closeness' or `distance'
 between spaces are indispensable. However there has been 
  no established mathematical theory so far which deals with these concepts 
  and which can be applied to spacetime physics\footnote{
  As a concept  which one should recall in this context, 
  there is  the Gromov-Hausdorff distance $d_{GH}(X,Y)$ between 
 two compact metric spaces~\cite{GR,PE}. Though it plays a central 
  role  in the convergence theory of 
 Riemannian geometry, its abstract nature may be   a big  obstacle  
 for its effective application to spacetime physics.}. 
  Here we will study  the averaging problem in cosmology 
 as {\it the problem of defining a distance between spaces}. 
 
 For this purpose  we would like to 
 focus on the {\it spectral representation} of spatial  
 geometry~\cite{MS-spectral}  as a promising attempt in this 
 direction.  The basic idea of the spectral representation is simple:  
 We utilize the `sound' of a space to characterize  the 
 geometrical structures of the space. This idea immediately reminds us 
  of a famous problem in mathematics, 
 `Can one hear the shape of a drum?'~\cite{MK} In imitation of this 
 phrase, we can state that 
 `Let us hear the shape of the universe!'~\cite{MS-spectral}

\section{The spectral distance}

 We should  materialize  the idea of `hearing the shape of the universe'
  in a definite form. For definiteness 
 we confine ourselves to $(D-1)$-dimensional Riemannian manifolds that are 
 spatial (metric signature $(+, \cdots, +)$) 
 and  compact without boundaries.  Let $Riem$ denote this class of 
 Riemannian manifolds.    
Now we set up the eigenvalue problem of the Laplace-Beltrami operator, 
 $\Delta f =-\lambda f$.
Then we obtain the {\it spectra}, viz. 
the set of eigenvalues (numbered in increasing order),   
$\{\lambda_n \}_{n=0}^\infty$.  
We note that, on dimensional grounds, the lower (higher) spectrum 
corresponds to the larger (smaller) scale behavior of geometry.

Suppose we want to compare two geometries $\cal G$ and ${\cal G}'$. 
Our strategy is hence to  compare 
the spectra $\{\lambda_n \}_{n=1}^N$ for $\cal G$
 with  the spectra $\{\lambda'_n \}_{n=1}^N$ for ${\cal G}'$. 
However,   taking a  difference $\lambda'_n - \lambda_n$ simply  
is not appropriate 
for our purpose: The simple difference  $\lambda'_n - \lambda_n$
would in general count the difference in the higher spectra 
  (corresponding to the smaller scale behavior of geometry) 
with more weight. In spacetime physics, however,  
the difference in the larger scale behavior of geometry is  
of more importance than the one  in the 
smaller scale behavior of  geometry. 
(This is the precise description  of the `spacetime foam picture' 
and the scale-dependent topology~\cite{MS-scale,MS-spectral}.) 
In addition,   the difference $\lambda'_n - \lambda_n$  has a 
physical dimension $[{\rm Length}^{-2}]$, which is not very 
comfortable either. Hence, we  should rather take the ratio 
$\frac{\lambda'_n}{\lambda_n}$; Then 
  the difference $\delta \lambda_n:= \lambda'_n- \lambda_n$ 
in the lower spectrum is counted with more weight as 
$\frac{\lambda'_n}{\lambda_n}=1+\frac{\delta \lambda_n}{\lambda_n}$.

Now  a measure of closeness $d_N({\cal G},{\cal G}')$ 
 between two geometries  $\cal G$ and ${\cal G}'$ can be introduced 
 by comparing the spectra 
$\{\lambda_n \}_{n=1}^N$ (for $\cal G$) with  
$\{\lambda'_n \}_{n=1}^N$ (for ${\cal G}'$) as   
\begin{equation}
  d_N ({\cal G},{\cal G}')= \sum_{n=1}^N {\cal F} 
  \left( \lambda'_n/ \lambda_n \right)\ \ .  
\label{eq:d_N_general} 
\end{equation}
Here the zero mode 
$\lambda_0=\lambda'_0=0$  is not included in the summation,  and $N$ is the  
cut-off number which can be treated as a  running parameter.  
The function ${\cal F}(x)$ ($x>0$) is a suitably chosen function 
which satisfies    ${\cal F} \geq 0$,  ${\cal F}(1)=0$,   
  ${\cal F}(1/x)={\cal F}(x)$,   and 
  ${\cal F}(y)>{\cal F}(x)$ if  $y > x \geq 1$. 
We also note that the cut-off number $N$ characterizes  up to which scale 
two geometries $\cal G$ and ${\cal G}'$ are compared. 
In this way,  $d_N ({\cal G},{\cal G}')$ 
is suitable for the scale-dependent description of the geometry. 

  At this stage, some comments may be appropriate on 
  the spectral representation  in general.  
  It is true that the spectra can be explicitly calculated only for 
  restricted cases. However, still there are several advantages for the 
  spectral representation.  
   First, the concept of the spectra itself is very clear. This is important 
   for practical applications in physics.
   Second, even when the exact spectra themselves are not known explicitly, 
   the perturbation analysis 
    gives us  important information on the spectra. 
    For instance, one can investigate the perturbed spectra around 
    some  well-understood spectra, just like one investigates the perturbed 
    metric around the Minkowski metric.  
    Third, in spacetime physics, 
   the lower spectra are more important than the higher spectra, since the 
   former spectra reflect the large scale structure of the universe. Thus, 
   even a few lower-lying spectra, which are in general easier to compute than 
   the higher spectra, carry important information.  
   (For more details, see Ref.~\cite{MS-spectral}.)

We note that the property 
${\cal F}\left( \lambda'_n/ \lambda_n \right)\rightarrow 0$ as 
$n \rightarrow \infty$ is 
required for the convergence of $d_N$ as $N \rightarrow \infty$.
Thus, it  follows  that  
$\l'_n/\l_n\rightarrow 1$ as $n \rightarrow \infty$ should hold 
 for convergence. 
Combined with  the Weyl's asymptotic formula~\cite{MK,CH}, it means that 
, in the $N \rightarrow \infty$ limit, 
the dimension and the volume of $\cal G$ and ${\cal G}'$ should be same 
in order to give a finite $d_N ({\cal G},{\cal G}')$ as 
$N \rightarrow \infty$~\cite{MS-spectral}. 
When $N$ is kept finite as most of the cases we consider, these conditions 
need not necessarily to be satisfied for  a finite $d_N$.

In order to utilize the measure $d_N ({\cal G},{\cal G}')$ efficiently, 
it is desirable that $d_N ({\cal G},{\cal G}')$  satisfies
 the axioms of distance, or at least some modified version 
of them:
\begin{description}
\item[(I)] {\it Positivity}: $d_N({\cal G}, {\cal G}')\geq 0$,  and 
      $d_N({\cal G}, {\cal G}')=0$ $\Leftrightarrow$ 
       ${\cal G} \sim {\cal G}'$, where $\sim$ means 
       equivalent in the sense of isospectral manifolds~\cite{MK,CH},  
\item[(II)] {\it Symmetry}: 
        $d_N({\cal G}, {\cal G}')=d_N({\cal G}', {\cal G})$,  
\item[(III)] {\it Triangle Inequality}:  
$d_N ({\cal G}, {\cal G}')+d_N ({\cal G}', {\cal G}'') \geq 
d_N ({\cal G}, {\cal G}'')$.  
\end{description}

Among several possibilities for the choice of ${\cal F}(x)$, there is one 
very important choice:
\begin{description}
\item[(a)] 
${\cal F}_a(x)=\frac{1}{2} \ln \frac{1}{2}(\sqrt{x}+1/\sqrt{x}) $.\\
 Then Eq.(\ref{eq:d_N_general}) becomes~\cite{MS-spectral}  
\begin{equation}
d_N ({\cal G},{\cal G}')
=\frac{1}{2} \sum_{n=1}^N \ln \frac{1}{2}
\left(
\sqrt{\frac{\lambda'_n}{\lambda_n}}
+\sqrt{\frac{\lambda_n}{\lambda'_n}}
\right)\ \ .
\label{eq:d_N} 
\end{equation}
It is notable that this form for $d_N$ 
can be related to the reduced density matrix
 element in quantum cosmology under some circumstances~\cite{MS-spectral}. 
 Namely, a long (short) spectral distance $d_N({\cal G}, {\cal G}')$ can be 
 interpreted as a strong (weak) quantum decoherence 
 between $\cal G$ and ${\cal G}'$ for some cases in  quantum cosmology.
This interpretation of $d_N$  gives one motivation for the choice of 
${\cal F}_a(x)$. 

The measure of closeness $d_N$ defined in Eq.(\ref{eq:d_N}) 
satisfies {\bf (I)} and {\bf (II)} of the distance axioms, but it does not 
satisfy the triangle inequality {\bf (III)}~\cite{MS-spectral}.
Significantly enough, however, the failure of the triangle inequality turns out 
to be only a mild one since  
a universal constant $c (>0)$ can be chosen such that 
$d'_N({\cal G},{\cal G}'):=d_N({\cal G},{\cal G}')+c$ 
recovers  the triangle inequality~\cite{MS-JGRG,MS-space}. 
Here $c$ is universal in the sense that $c$ can be chosen independent of 
${\cal G}$, ${\cal G}'$ and ${\cal G}''$ although it depends 
on $N$. This fact leads to a significant consequence 
below: $d_N$ and its  modification $\bar{d}_N$ (see below) are 
closely related to each other, which helps us  reveal the 
  nice properties of $d_N$.   
\end{description}

There is another important choice for ${\cal F}(x)$:
\begin{description}
\item[(b)]  ${\cal F}_b(x):=\frac{1}{2}\ln\max(\sqrt{x},1/\sqrt{x})$. \\
This  is a slight modification of ${\cal F}_a$. 
In this case, Eq.(\ref{eq:d_N_general}) becomes 
\begin{equation}
\bar{d}_N({\cal G}, {\cal G}')
=\frac{1}{2}\sum_{n=1}^{N}
\ln\max\left(\sqrt{\frac{\lambda_n'}{\lambda_n}},
           \sqrt{\frac{\lambda_n}{\lambda_n'}} \right) \ \ . 
\label{eq:d_N_bar}
\end{equation}
Note that $\bar{d}_N$ satisfies {\bf (I)}-{\bf (III)}, so that 
it {\it is} a distance. 
\end{description}

  Now, each  measure of closeness introduced above has its own advantage: 
$d_N$ in {\bf (a)} has an analytically neat form and it 
can be related to the quantum decoherence between $\cal G$ and ${\cal G}'$ 
in the context of quantum cosmology; However, it does not satisfy the triangle 
inequality. On the other hand,  the measure $\bar{d}_N$ in {\bf (b)} is 
a distance in a rigorous sense, although  its form is not very convenient 
for practical applications (it contains $max$). Quite surprisingly, 
it turns out that $d_N$ and 
$\bar{d}_N$ are deeply related to each other. 
To discuss this property, we 
introduce  an $r$-ball centered at ${\cal G}$ defined by $d_N$ 
 in  Eq.(\ref{eq:d_N}),  
\[
B({\cal G}, r; d_N )
:=\{{\cal G}'\in Riem/_\sim| 
   d_N \left({\cal G}, {\cal G}' \right)<r  \}\ \ .
\]
Here  $\sim$ indicates 
the identification of isospectral manifolds. In the same manner,  
we also introduce    an $r$-ball centered at ${\cal G}$ defined by $\bar{d}_N$, 
$B({\cal G}, r; \bar{d}_N )$. 

Now we can show that~\cite{MS-space} 
\begin{description}
\item{\bf{Theorem 1}} \\
The set of balls 
$\{ B({\cal G}, r; d_N)|\ {\cal G}\in Riem/_\sim ,\ r>0 \}$
and the set of balls 
$\{ B({\cal G}, r; \bar{d}_N)|\ {\cal G}\in Riem/_\sim,
 \ r>0 \}$ generate  the same topology on $Riem/_\sim$.
\end{description}

For the proof of {\it Theorem 1},  first we should show that 
the set of all balls 
$\{ B({\cal G}, r; d_N)|\ {\cal G}\in Riem/_\sim ,\ r>0 \}$ can 
actually define a topology (let us call it ``$d_N$-topology''), viz.
the set of all balls can be a basis of open sets. This property is 
far from trivial, because of the failure of the triangle inequality for 
$d_N$.\footnote{
On the other hand, a similar set of balls defined by $\bar{d}_N$ 
can define a topology (let us call it $\bar{d}_N$-topology), since 
$\bar{d}_N$ is a distance.} 
Next, we need to show that any ball defined by $d_N$ (resp. $\bar{d}_N$) 
is an open set in $\bar{d}_N$-topology 
(resp. $d_N$-topology)~\cite{MS-space}.

From {\it Theorem 1}, we immediately obtain 
\begin{description}
\item{\ \ \ \bf{Corollary}} \\
The space 
$
{\cal S}_N^o
:=\left(Riem, d_N \right)/_\sim
$
 is a metrizable space. The distance 
function for metrization is provided by $\bar{d}_N$.
\end{description}
 
Hence we can  extend ${\cal S}_N^o$ to its completion\footnote{
It is desirable to investigate the structure of ${\cal S}_N$ 
intensively as a purely mathematical object.},  
${\cal S}_N$. 
Due to {\it Theorem 1} and its {\it Corollary}, 
it is justified to treat $d_N$ as a distance and 
to  regard ${\cal S}_N$ as a metric space, provided that    
we  resort to the distance function $\bar{d}_N$ whenever 
the triangle inequality  is needed in the arguments.  

We can also show that ${\cal S}_N$ has  several other 
desirable  properties~\cite{MS-space}:
\begin{description}
\item{\bf Theorem 2} \\
  The space  ${\cal S}_N$ is paracompact.
\end{description}
\begin{description}
\item{\ \ \ \bf Corollary} \\
There exists partition of unity subject to 
 any open covering of ${\cal S}_N$.
\end{description} 
\begin{description}
\item{\bf Theorem 3}\\
The space ${\cal S}_N$ is locally compact.

Due to this property, we can construct an integral over 
${\cal S}_N$~\cite{LAN}, which is essential to consider, e.g.,  
 probability distributions over ${\cal S}_N$. 
\item{\ \ \ \bf Corollary}\\ 
If a sequence of continuous functions on ${\cal S}_N$, 
$\{ f_n \}_{n=1}^\infty$, 
pointwise converges  to a function $f_\infty$, then 
$f_\infty$ is continuos on a dense subset of ${\cal S}_N$. 
\end{description} 
\begin{description}
\item{\bf Theorem 4} \\
The space ${\cal S}_N$ satisfies the second countability axiom.
\end{description} 

These properties of ${\cal S}_N$ suggest that 
the space ${\cal S}_N$ can serve as a basic arena for 
spacetime physics. 
From now on  we call $d_N$ in Eq.(\ref{eq:d_N_general})  
  (the form of $d_N$ in Eq.(\ref{eq:d_N}) in particular) 
 a {\it spectral distance} for brevity. 

\section{Model-fitting procedure in cosmology}

Now let us come back to the averaging problem in cosmology. 
Regarding  this problem, there are several underling issues as follows:
\begin{description}
\item[(1)] How to  select out a time-slicing  for a given spacetime (`reality'),
which in general would  possess no symmetry. 
\end{description}
Furthermore, 
\begin{description}
\item[(2)] How to incorporate the spatial diffeomorphism invariance, 
\item[(3)] How to incorporate the scale-dependent aspects  of the 
geometrical structures, and
\item[(4)] How to incorporate the apparatus dependence of 
the observed information
\end{description}
 to the averaging procedure of  geometry.

Considering its several desirable properties,   
the spectral representation  seems to  
serve as a suitable `language' for formulating 
and analyzing these issues. In particular the space ${\cal S}_N$ introduced 
in the previous section provides  an appropriate platform for these 
discussions. 

As a demonstration, let us give a rough sketch of the mapping procedure from 
reality to a model in terms of the spectral representation. 

We here consider how to assign a model spacetime to a given 
 spacetime  (`reality'). 
We first fix notations. 
\begin{description}
\item{$(1^\circ)$}
  We consider a portion of a spacetime 
  $({\cal M}, g)$ bounded by   two non-intersecting spatial sections 
  $\Sigma_0$ and $\Sigma_1$ of 
  $({\cal M}, g)$. Let us denote this portion as 
  $({\cal M}, g)_{(0,1)}$.
\item{$(2^\circ)$}
  Let $Slice_o ({\cal M}, g)_{(0,1)}$ be a set of all possible 
  time-slicings of  $({\cal M}, g)_{(0,1)}$. 
\item{$(3^\circ)$}
 Hence,  a slice $s \in Slice_o ({\cal M}, g)_{(0,1)}$ can be identified with 
    a parameterized set of spatial geometries 
  $\{ \left(\Sigma, h(\beta) \right) \}_{0\leq \beta \leq 1}$. 
Let $({\cal M}, g)_{(0,1),s}$ denote this set  for brevity. 
\item{$(4^\circ)$}
Let $\{models \}$ be a set of model spacetimes,  
bounded by   two non-intersecting spatial sections 
$\Sigma'_0$ and $\Sigma'_1$ with 
a particular time-slicing, $({\cal M}', g')_{(0,1),s'}$. 
Here, we  distinguish  between 
the identical spacetimes  $({\cal M}', g')$ with 
different choices of two  non-intersecting spatial sections 
($\Sigma'_0$ and $\Sigma'_1$) and/or 
different choices of a time-slicing ($s'$).  
 \end{description}
It is notable that the metric-space structure of ${\cal S}_N$ induces 
the  same structure on $Slice_o ({\cal M}, g)_{(0,1)}$ also:   
 Let 
 $s_1:=\{ \left(\Sigma, h_1(\beta) \right) \}_{0\leq \beta \leq 1}$ 
 and 
 $s_2:=\{ \left(\Sigma, h_2(\beta) \right) \}_{0\leq \beta \leq 1}$ 
  are any elements in $Slice_o ({\cal M}, g)_{(0,1)}$. Then,  
  we can  define 
\begin{equation}   
D_N(s_1,s_2):=
\int_0^1 \left[
   d_N
   \left( \left(\Sigma_1, h_1(\beta)\right), 
           \left(\Sigma_2, h_2(\beta)\right)
   \right) \right] d\mu(\beta)\ \ , 
\end{equation}    
where $\mu(\beta)$ is a positive-definite measure. 
 Clearly $Slice_o ({\cal M}, g)_{(0,1)}$ with $D_N$  
 becomes a metrizable space reflecting the same property of 
 ${\cal S}_N$.  Thus, we can consider its completion,  
 $Slice({\cal M}, g)_{(0,1)}$.

Now we describe the procedure of assigning  a model spacetime 
to reality. 
\begin{description}
\item{[1]} {\it The choice of time-slicing}

   Let us choose and fix one model spacetime with a particular time-slicing 
   $({\cal M}', g')_{(0,1),s'}$. We can select 
   the most suitable time-slicing of $({\cal M}, g)_{(0,1)}$
    w.r.t. (with respect to) 
    the model $({\cal M}', g')_{(0,1),s'}$ as follows:  
    For each parameter $\beta$ ($0<\beta<1$), 
    the closeness between  the slice $\left(\Sigma, h(\beta)\right)$ 
    in $({\cal M}, g)_{(0,1),s}$ and the slice 
    $\left(\Sigma', h'(\beta)\right)$ in $({\cal M}', g')_{(0,1),s'}$  can be 
    measured by the spectral distance 
  $d_{A,\Lambda}
 \left(\left(\Sigma,  h(\beta)\right), \left(\Sigma', h'(\beta)\right)\right)$.
   Here the suffixes $A$ and $\Lambda$ indicate, respectively,  
   the elliptic operator (we here consider the Laplacian for simplicity) 
   and the cut-off number (viz. $N$ in the previous section) for defining 
   the spectral distance. Physically, $A$ and $\Lambda$ symbolize 
   the observational apparatus and the cut-off scale, respectively.

   Now we can select out 
     the most suitable time-slicing of $({\cal M}, g)_{(0,1)}$
    w.r.t.  
    the model $({\cal M}', g')_{(0,1),s'}$  as 
\vskip .5cm
\begin{minipage}{130mm} 
{\large {\bf \hskip 1cm  Select $ s_0 \in Slice({\cal M}, g)_{(0,1)}$ s.t. \\
  \begin{eqnarray*}
 && D_{A,\Lambda}
      \left(({\cal M}, g)_{(0,1),s}, ({\cal M}', g')_{(0,1),s'}\right) \\
 &&\qquad  \qquad  := \int_0^1 \left[
   d_{A,\Lambda}
   \left( \left(\Sigma, h(\beta)\right), \left(\Sigma', h'(\beta)\right)
   \right) \right] d\mu(\beta)
   \end{eqnarray*}
  \hskip 1cm  gives the minimum.}}
\end{minipage} 
  \vskip .5cm
On account of the property that  $D_{A,\Lambda}$ is bounded from below 
along with the completeness of $Slice({\cal M}, g)_{(0,1)}$,  
  some time-slicing $s_0$ of $({\cal M}, g)_{(0,1)}$ is 
 selected out  w.r.t. the 
 model $({\cal M}', g')_{(0,1),s'}$, $A$ (apparatus) and  
 $\Lambda$ (scale)\footnote{
 To be more precise, there can be more than one slicings that satisfy the 
 condition. Furthermore, $s_0$ can be a limit point of 
 $Slice_o({\cal M}, g)_{(0,1)}$, viz. 
 $s_0 \in Slice({\cal M}, g)_{(0,1)}\setminus Slice_o({\cal M}, g)_{(0,1)}$. 
 In such a case, one would judge that 
  $({\cal M}', g')_{(0,1),s'}$ is not an appropriate model for 
  $({\cal M}, g)_{(0,1),s}$. In any case, it is desirable to investigate 
 the mathematical structure of $Slice({\cal M}, g)_{(0,1)}$ in more detail. 
 }.    
  
  \item{[2]} {\it Assignment of a model to `reality'}  
 
  We can continue  the same procedure for every model spacetime 
  $\in \{models \}$
  to choose the best-fitted model $({\cal M}^*, g^*)_{(0,1),s^*}$ and, w.r.t. 
  it,  the time-slicing $s^*_0$ 
  of $({\cal M}, g)_{(0,1)}$. Then one can regard 
   $({\cal M}_*, g_*)_{(0,1),s^*}$ to be the cosmological counterpart of 
   $({\cal M}, g)_{(0,1),s^*_0}$ w.r.t. $(A, \Lambda)$. 
   In this way the spectral 
   representation naturally leads us to the concept of {\it apparatus- and 
   scale-dependent effective evolution of the universe}. 
\end{description}

\section{Example:(2+1)-dimensional flat spacetimes}
As an illustration for the procedure in the previous section, 
let us consider a simple example.  We 
choose as  `reality' the simplest (2+1)-dimensional flat 
spacetime with  topology $ T^2 \times \mbox{\boldmath $R$} $:  
We can  construct such a spacetime  
 from $\mbox{\boldmath $R$}^3$ by the identification in space,  
 $(x+m,\ y+n) \sim (x,\ y)$, where $m,n \in \mbox{\boldmath $Z$}$. 
 (Here, $(x,y,t)$ is the standard coordinates for $\mbox{\boldmath $R$}^3$.)
 We can imagine this spacetime as 
 a static spacetime with a spatial section being 
 a regular 2-torus (a torus constructed from a unit square by gluing 
 the edges facing each other), if $t=const$ slicing is employed.   
Now, let   $({\cal M}, g )_{(0,1)}$ be a portion of the spacetime 
 defined by  $0 \leq t \leq 1$.  
 Then $Slice_o({\cal M}, g )_{(0,1)}$ denotes a set of 
 all slices for the present $({\cal M}, g )_{(0,1)}$, and  
$Slice({\cal M}, g )_{(0,1)}$ is its completion.
 
As   a set of model spacetimes, $\{ models \}$,  we take 
a set of all (2+1)-dimensional flat spacetimes of  topology 
 $ T^2 \times \mbox{\boldmath $R$} $ with particular slices; 
 For each model, 
 a particular  time-slicing  is employed 
 by which  the line-element  is 
 represented as  
\[
ds^2= -dt^2 + h_{ab} d\xi^a d\xi^b\ \ \ ,
\]
where
\[
h_{ab}  ={ V \over  \tau^2 }
\pmatrix{ 1     &  \tau^1  \cr
         \tau^1 &  |\tau|^2 \cr}\ \ \ . 
\]
Here 
 $(\tau^1, \tau^2)$ are the Teichm\"uller parameters of a 2-torus, 
 and $\tau := \tau^1 + i \tau^2$, 
$\tau^2 >0$;   
$(\tau^1, \tau^2)$ and $V (>0)$ are functions of $t$ only; 
The periodicity in the coordinates $\xi^1$ and $\xi^2$ with 
period 1 are understood.   
We note that $(\tau^1,\tau^2)$ represent the shape of a 
parallelogram\footnote{
In the present parametrization, the coordinates of four vetices of 
the parallelogram $OACB$ are $O=(0,0)$, 
$A=(\tau^1/\sqrt{\tau^2}, \sqrt{\tau^2})$, 
$B=(1/\sqrt{\tau^2}, 0)$ and 
$C=(\frac{1+\tau^1}{\sqrt{\tau^2}},\sqrt{\tau^2} )$~\cite{MS-(2+1)}.  
} which forms the 2-torus by the edge-gluing; $V$ 
 represents  the 2-volume of the 2-torus~\cite{MS-(2+1)}. 

The functional forms for $\tau^1$, $\tau^2$ and $V$ are not arbitrary;  
The evolutions of $\tau^1$, $\tau^2$ and $V$ w.r.t. $t$ are determined  by 
a simple constrained Hamiltonian system~\cite{HN} \\
$\{(\tau^1, p_1),  \ (\tau^2, p_2),  \ (V,\sigma);  \ H \simeq 0\}$.
 Thus, in this example, $\{models\}$ is parameterized by 
 distinct initial conditions for the Hamiltonian system. In other words, 
  4 parameters are required in principle to characterize each model in  
  $\{models\}$.  

Now, take one model in $\{ models \}$, and consider its portion 
characterized by 
$0 \leq t \leq 1$. This portion of the model corresponds to 
$({\cal M}', g' )_{(0,1),s'}$ in the previous section. 
We easily get the spectra for  
each time-slice $\Sigma'$ of $({\cal M}', g' )_{(0,1),s'}$:  
The Laplacian in this case becomes 
$\Delta=h^{ab}\frac{\partial}{\partial \xi^a}$
$\frac{\partial}{\partial \xi^b}$; 
The normalized eigenfunctions of the Laplacian are  
\[
f_{n_1 n_2} (\xi^1, \xi^2) 
= \exp (i2\pi n_1 \xi^1)\cdot  \exp (i2\pi n_2 \xi^2)
\]
with the  spectra  
\begin{eqnarray}
\lambda'_{n_1 n_2}
   &=& \frac{4\pi^2}{V\tau^2}|n_2-\tau n_1|^2 \nonumber \\
   &=& \frac{4\pi^2}{V\tau^2}(|\tau|^2 n_1^2-2\tau^1 n_1 n_2 + n_2^2) \ \ \  
 \ \ \ (n_1, n_2  \in  {\bf Z})\ \ \ .
\label{eq:spectra-model}
\end{eqnarray}
On the other hand, the  
`reality' $({\cal M}, g )_{(0,1)}$ is identical with 
an element of $\{models\}$ when $t=const$ slicing $s_c$ is employed; viz. 
$({\cal M}, g )_{(0,1),s_c}$ is identical with 
the model characterized by  $\tau^1\equiv 0$, $\tau^2\equiv 1$ and 
$V\equiv 1$. Then, for every spatial section
 $\Sigma$ of $({\cal M}, g )_{(0,1),s_c}$, 
 the spectra become
\begin{equation}
\lambda_{n_1 n_2}= 4\pi^2 
 ( n_1^2 + n_2^2)\ \ \ 
 \ \ \ (n_1, n_2  \in  {\bf Z})\ \ \ .
\label{eq:spectra-reality}
\end{equation}

We can measure the spectral distance with the help of 
Eqs.(\ref{eq:spectra-model}) and (\ref{eq:spectra-reality})
   $d_N
   ( \left(\Sigma, h(t)\right), $$\left(\Sigma', h'(t)\right))$. 
   It is obvious that 
   $D_N
      (({\cal M}, g)_{(0,1),s}, $$({\cal M}', g')_{(0,1),s'})$  
  gives the absolute minimum, 0,  only when the model  
  $({\cal M}', g')_{(0,1),s'}$ 
  is the one characterized by  $\tau^1\equiv 0$, $\tau^2\equiv 1$ and 
$V\equiv 1$, and the slicing of the `reality' is $s=s_c$.

\section{Dynamics of spectra}

We have established   the spectral distance, which 
provides  a  basis for comparing  the real spatial geometry 
with a model spatial geometry. 
We can now investigate the spectral distance between 
`reality' and a model as a function of time,
 which serves  as the quantitative  analysis of  the 
influence of the averaging procedure on the effective dynamics of the 
universe.  
Here we see the usefulness of the spectral distance: On one hand 
it has a nice mathematical properties, and on the other hand, it can be 
handled explicitly. 
Thus, we now need dynamical equations for the spectra.   

We first prepare concise notations for  specific  integrals that appear 
frequently below.  
 Let  $A(\cdot)$ and $A_{ab}(\cdot)$ be 
  any function and any symmetric tensor field, respectively, 
  defined on a spatial section $\Sigma$. 
 Let $\{ f_n \}_{n=0}^\infty$ be the eigenfunctions of the Laplacian. 
 Then we define  
\begin{eqnarray*}
\langle A  \rangle_{mn}:&=&\int_{\Sigma} f_n\  A(x)\  f_m \ \ , \ \ 
\langle A  \rangle_{n}:=\langle A  \rangle_{nn} \ \ , \\
\langle A_{ab} \rangle_{mn} :&=& 
    \frac{1}{\sqrt{\lambda_m \lambda_n}} 
     \int_{\Sigma} \partial^a f_m\  A_{ab}(x)\   \partial^b f_n \ \ .
\end{eqnarray*}

In order to derive the spectral evolution equations, 
 we first recall  a basic result of the perturbation theory 
 (``Fermi's golden rule'')
\begin{equation}
\d \l_n = - \<\d \Delta {\>_{}}_{n}  \ \ . 
\label{eq:dlambda}             
\end{equation}
Noting 
$\Delta f= \frac{1}{\sqrt{h}}(\sqrt{h} \ h^{ab}\partial_b f),_{a}$, it is 
straightforward to get 
\begin{equation}
\<\d \Delta {\>_{}}_{n} = \<\overline{\d h}_{ab}{\>_{}}_{n}\l_n +   
                \frac{1}{2}\<h\cdot \d h {\>_{}}_{n}\l_n \ \ , 
\label{eq:dDelta}
\end{equation}
where $h\cdot \d h:=h^{ab}{\d h}_{ab} $ and 
$\overline{\d h}_{ab}:={\d h}_{ab}-\frac{1}{2} h\cdot \d h\  h_{ab}$. 
Combining Eq.(\ref{eq:dDelta}) with Eq.(\ref{eq:dlambda}), simple manipulations
lead to a formula~\cite{MS-dynamics}
\begin{equation}
\d \l_n = -\<\d h_{ab}{\>_{}}_{n}\l_n
 +\frac{1}{4}\<\Delta(h\cdot \d h){\>_{}}_{n}\ \ .
\label{eq:dlambda2}
\end{equation}
Now we identify $\d h_{ab}$ in Eq.(\ref{eq:dlambda2})
  with the time-derivative of the spatial metric, 
$\dot{h}_{ab}$, w.r.t. the time-slicing: $\d h_{ab}$ should be   
 replaced by $\dot{h}_{ab}= 2 N K_{ab}+ 2 D_{(a}N_{b)}$. Here 
 $N$ and $N_a$ are the lapse function and the shift vector, respectively; 
 $K_{ab}$ is the extrinsic curvature and $K:=K_a^{\ a}$. 
 After some manipulations~\cite{MS-dynamics}, 
 we finally reach the basic formula for the 
 spectral evolution, 
\begin{equation}
\dot{\lambda}_n=-2\langle N K_{ab} \rangle_n \lambda_n +
         \frac{1}{2}\langle \Delta (NK) \rangle_n \ \ .  
\label{eq:dynamics}         
\end{equation}
We note that 
the shift vector $N_a$ does not appear  in the final result,
 Eq.(\ref{eq:dynamics}).  
This result comes from the fact that   
the spectra are spatial diffeomorphism invariant quantities. 

For simplicity, let us set $N \equiv 1$. Then we get 
\begin{equation}
\dot{\lambda}_n=-\frac{2}{D-1} \langle  K  \rangle_n \lambda_n
         + \frac{D-3}{2(D-1)}\langle \Delta K \rangle_n 
         -2 \langle  \epsilon_{ab}  \rangle_n \lambda_n \ \ , 
\label{eq:dynamics2}         
\end{equation}
where $\epsilon_{ab}:= K_{ab}- \frac{1}{D-1}K h_{ab}$, 
 and  $D$ is the spacetime dimension ($D=4$ in the ordinary case).
  
  It is not our present aim to go into the detailed analysis on 
   the dynamics of the spectra, 
   which will be done elsewhere~\cite{MS-dynamics}.
    Here we only discuss a simple example as a demonstration of the 
  usefulness of the spectral scheme: Let us investigate 
   the scale-dependence of the effective Hubble parameter $H_{eff}$. 
   
  Suppose the spacetime is close to the closed Friedman-Robertson-Walker 
  universe. In this case, $\Delta K$  and $\epsilon_{ab}$   are regarded as 
  small. (More precisely, $\langle \Delta K \rangle_{mn}$ 
   and  $\langle \epsilon_{ab} \rangle_{mn}$ 
   are small compared to  $\langle K \rangle_{mn} \sqrt{\lambda_m \lambda_n}$ 
   and $\langle K \rangle_{mn}$,  
  respectively ($m,n=1,2,\cdots$).)  
  Then Eq.(\ref{eq:dynamics2}) can be written as (we set $D=4$) 
\begin{equation}
\dot{\lambda}_n = -\frac{2}{3} \langle K \rangle_n
\left[ 
1 -\frac{1}{4}
   \frac{\langle \Delta K \rangle_n}{\langle K \rangle_n \lambda_n}
  + 3\frac{\langle \epsilon_{ab} \rangle_n}{\langle K \rangle_n}
   \right] \lambda_n\ \ ,
\label{eq:dynamics3}         
\end{equation}
hence
\begin{equation}
(H_{eff})_n=
\frac{1}{3} \langle K \rangle_n
\left[ 
1 -\frac{1}{4}
   \frac{\langle \Delta K \rangle_n}{\langle K \rangle_n \lambda_n}
  + 3\frac{\langle \epsilon_{ab} \rangle_n}{\langle K \rangle_n}
   \right]\ \ .
\label{eq:Heff}         
\end{equation}
The last two terms in the bracket describe  the influence of  
inhomogeneity and anisotropy  on  $H_{eff}$ {\it at scale} $\Lambda$. 
Here we note once more that the spectral representation naturally describes 
the apparatus- and the scale-dependent picture of the universe (in the present 
example, $(H_{eff})_n$).

\section{Spacetime picture from the viewpoint of the spectral representation}

We have discussed the spectral representation of geometrical structures
in connection with the averaging problem in cosmology. 
In particular we have  introduced   ${\cal S}_N$, 
the space of all spaces 
equipped with the spectral distance, and have shown that 
${\cal S}_N$ possesses several desirable properties 
as a basic arena for spacetime physics.
    
     We have  sketched the model-fitting procedure 
     in the framework  of the spectral representation, and have also 
     suggested how to analyze the dynamical aspects of the averaging procedure 
     within  this framework.  
     These arguments imply  that 
     the spectral scheme  seems to be suitable for the analysis of 
     the averaging problem. In fact, it naturally describes 
     the apparatus- and scale-dependent effective evolution of 
     the universe.  
     
   Finally, let us briefly discuss 
   how spaces look like from the viewpoint of the spectral representation.
    
   One may imagine the whole of the   
   geometrical information of a space as a collection of 
   all spectra such as 
\[
Space = \bigcup_i \large({\cal D}_i, \{\lambda^{(i)}_n \}_{n=0}^\infty, 
\{f^{(i)}_n \}_{n=0}^\infty \large)\ \ ,
\]
  where ${\cal D}_i$ denotes an elliptic operator and  
  $\{\lambda^{(i)}_n \}_{n=0}^\infty$ and $\{f^{(i)}_n \}_{n=0}^\infty$ 
   are  its spectra and the eigenfunctions, respectively.
    The index $i$ runs over all possible elliptic operators. 
  A single  observation is related to 
    a subclass of elliptic operators corresponding to 
  the observational apparatus. Thus  we get  
   only a small portion of the whole geometrical 
  information of the space  by  a single   observation.
  Such incomplete information may not 
   be enough to determine geometry  uniquely.  
   Only one has to do then is  to 
   conduct   other kinds of observation corresponding to other types of  
   elliptic operators  in order to get    further information on geometry.
   (This is the physical interpretation of `isospectral manifolds', viz. 
   non-isometric manifolds with the identical spectra of the Laplacian.) 
    It is also tempting to regard  the  spectral information  
   more fundamental than the concept of Riemannian manifolds. 
   Further investigations are needed 
   to judge to what extent  such a viewpoint of spacetime geometry is valid.




\begin{thebibliography}{99}
\bibitem{AVE}
For instance, G.F.R. Ellis, in {\sl ``Proceedings of the Tenth 
International Conference on General Relativity and Gravitation''}, 
edited by B. Bertotti, F. De Felice and A. Pascolini 
(Reidel, Dordrecht, 1984); H. Sato, {\it ibid}; 
T. Futamase, 
 Phys. Rev. Let. {\bf 61}, 2175 (1988); Mon. Not. R. astr. Soc. 
 {\bf 237}, 187 (1989).
\bibitem{AVE2} For a concise review of the averaging problem, see 
A. Krasi{\'n}ski, {\sl Inhomogeneous Cosmological Models} 
(Cambridge University Press, Cambridge, 1997), Chapter 8. See also  the 
references therein. 
\bibitem{GR}
M. Gromov,  J. Lafontaine,  P. Pansu,   
{\sl structures m{\'e}triques pour les vari{\'e}t{\'e}s riemannienness}    
(Cedic/Fernand Nathan, Paris,   1981).
\bibitem{PE}
P. Petersen,   {\sl Riemannian Geometry} 
  Springer-Verlag, New York,  1998),  Chapter 10.  
\bibitem{MS-spectral}
M. Seriu, Phys. Rev. D{\bf 53}, 6902 (1996).
\bibitem{MK}
M. Kac, Am. Math. Mon. {\bf 73}(4), 1 (1966). 
\bibitem{MS-scale}
M. Seriu, Phys. Let. B{\bf 319}, 74 (1993);
  Vistas in Astronomy {\bf 37}, 637 (1993).
\bibitem{CH} 
See e.g., I. Chavel,  {\sl Eigenvalues in Riemannian Geometry}   
(Academic Press, Orland,  1984).
\bibitem{MS-JGRG}
M. Seriu, in {\sl ``Proceedings of the 8th Workshop
 on General Relativity and Gravitation"}  
  (K.Oohara et. al. (eds.), Niigata University, 1999).  
\bibitem{MS-space}
M. Seriu, {\sl ``Space of spaces as a metric space"}, 
gr-qc/9908078, to appear in Comm. Math. Phys.  
\bibitem{LAN}
S. Lang, {\sl Real Analysis} (Addison-Wesley, Reading, 1969), Chapter 12. 
\bibitem{MS-(2+1)}
M. Seriu, Phys. Rev. D{\bf 53}, 1889 (1996). 
\bibitem{HN} 
A. Hosoya and K. Nakao, Prog. Theo. Phys. {\bf 84}, 739 (1990). 
\bibitem{MS-dynamics}
M. Seriu, {\sl ``Spectral evolution of the Universe"}, 
submitted for publication. 
\end{thebibliography}
\end{document}